\documentclass[mathleft]{an}
\usepackage{graphicx}
\usepackage{times}
\usepackage{natbib}
\bibpunct{(}{)}{;}{a}{}{,}
\begin{document}

\Pagespan{789}{}
\Yearpublication{2006}%
\Yearsubmission{2005}%
\Month{11}%
\Volume{999}%
\Issue{88}%

\title{RR Lyrae stars seen from space}

\author{E. Guggenberger\inst{1}\fnmsep\thanks{Corresponding author:
  \email{elisabeth.guggenberger@univie.ac.at}\newline}
}
\titlerunning{RR Lyrae stars seen from space}
\authorrunning{E. Guggenberger}
\institute{
Institut f\"ur Astronomie, Universit\"at Wien, T\"urkenschanzstr. 17, 1180 Vienna, Austria
}

\received{30 May 2005}
\accepted{11 Nov 2005}
\publonline{later}


\abstract{
RR Lyrae stars for a long time had the reputation of being rather simple pulsators, but the advent of high-precision space photometry has meanwhile changed this picture dramatically. This article summarizes the results obtained for two remarkable Blazhko RR~Lyrae stars and discusses how our view of RR Lyrae stars has changed since the availability of  ultra-precise satellite photometry as it is obtained by CoRoT and {\it Kepler}. Both stars, CoRoT 105288363 and V445~Lyrae, show a multitude of phenomena that were impossible to observe from the ground, either because of the small amplitude of the effect, or because uninterrupted long-term monitoring was required for a detection. Not only was it found that strong and irregular cycle-to-cycle changes of the Blazhko effect can occur, and that seemingly chaotic phenomena need to be accounted for when modeling the Blazhko effect, but also a rich spectrum of low-amplitude frequencies was detected in addition to the fundamental radial pusation in RRab stars. The so-called period doubling phenomenon, higher radial overtones and possibly also non-radial modes make RR Lyrae stars more multifaceted than previously thought. This article presents the various aspects of irregularity of the Blazhko effect, questioning its long-standing definition as a "periodic modulation", and also discusses the low-amplitude pulsation signatures that had been hidden in the noise of observations for centuries.}

\maketitle

\section{Simple or complex?}
Because of their large amplitudes (about 1 mag in V) that makes them easy to discover, RR Lyrae stars have been known since the end of the 19th century. It was found by \citet{schw} that they pulsate either in the fundamental radial mode (subtype RRab) or in the first radial overtone (subtype RRc). Double mode RR Lyrae (subtype RRd) were discovered in 1977 \citep{jer}. Because of their pure radial pulsation and their long history they were considered rather simple and well-studied stars. They are used as distance indicators and form the bottom of the cosmic distance ladder. Even the long-standing unsolved problem, the Blazhko effect \citep{blazhko}, which is known to affect about  40-50 \% of the RRab stars \citep{jur, ben, kol} was assumed in certain ways to be a ``well-behaved'' problem. It is usually defined as ``a periodic amplitude and/or phase modulation of the order of several ten to hundred times the pulsation period''. Therefore, only one Blazhko period and one modulation amplitude was assigned to each star, the cycles were expected to repeat exactly, and theoretical models linked the effect to rotation, either directly or indirectly, therefore requiring clock-work like repetition. But are RR Lyrae stars really so simple or ``well-behaved''?

\subsection{Some older indications of not so simple behaviour} 
When searching the literature, one can find several clues that the Blazhko effect might be more complex. There are discussions about various phenomena, such as possible secondary cycles, the ceasing of the effect in certain stars, and the subsequent reappearance of the modulation, as well as differing Blazhko periods when the same target was measured in different years. Those clues, however, never lead to a change in the wide-spread and simplified overall picture of a periodic Blazhko effect, even though they have been known for a long time. The oldest publication on a changing Blazhko effect in fact dates back to the 1950s, when \citet{bala} reported that the amplitude of the phase modulation of RW~Dra had dropped dramatically. Within 3 months, the amplitude of the variation in the O-C diagram had changed from 2h17m to only 1h. \\
Probably the most famous case of a complex Blazhko modulation is the prototype RR~Lyrae itself, which is known to have a 4-yr cycle in addition to the 40-d Blazhko period. This additional cycle is superimposed on the modulation and quenches the modulation amplitude periodically, so that the Blazhko effect became almost undetectable in the years 1963, 1967, 1971, and 1975 \citep{szeidl}.\\
A star in which two modulation periods have been unambiguously detected is CZ~Lac \citep{sodor}. The two periods lead to a beating in the Blazhko modulation of this star.

\section{CoRoT 105288363 and V445 Lyr}
The major problem about the older reports on complex Blazhko behaviour was the incomplete sampling of the data. Ground based data suffered from small as well as large gaps due to daylight and bad weather. It was therefore impossible to say whether a change in the Blazhko phenomenon or a change of the pulsation period had taken place slowly and continuously, or whether it had happened abruptly. A discrimination between slow evolutionary changes and sudden glitches was usually impossible. Also, it was often necessary to compare observations that were obtained with different instrumentations by different authors, and the reliability of the result based on the merged data was questionable. These problems are not present at all in the long time-series of uninterrupted ultra-precise space photometry. The first star, for which a changing Blazhko effect was observed in such a data set, was CoRoT 105288363. The time base of the observations was 145~d, so that a total of four complete Blazhko cycles could be observed without interruption, making it possible to observe the changes as they took place. Differences between the cycles were large enough to be visible even when looking at the raw data, and a sampling time of 512~s guaranteed complete coverage of all features in the the light curves. A detailed analysis of this data set was performed by \citet{gug11}. The second star of this kind, V445~Lyr, which was observed by {\it Kepler} turned out to be even more extreme, with even larger differences between the Blazhko cycles. The total time base of the data set that was published by \citet{gug12} was 588~d with eight complete Blazhko cycles covered.

\subsection{Irregular and multi-periodic modulation}
\begin{figure}
\includegraphics[width=82mm]{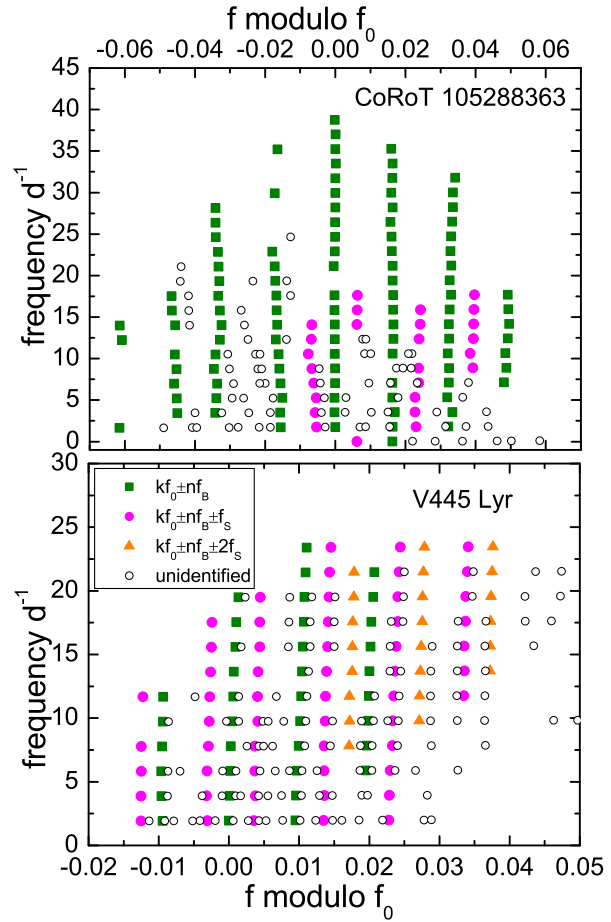} 
\caption{Echelle diagrams illustrating the frequency patterns in the Fourier spectra of CoRoT 105288363 (upper panel) and V445 Lyr (lower panel).  Green squares indicate the classical Blazhko multiplet, filled magenta circles indicate peaks belonging to the secondary multiplet, while open circles display the numerous close peaks that are caused by irregular variation. Filled triangles represent second order components of the secondary modulation, which could only be detected in V445 Lyr. Please note the different ranges on the x-axis which are due to the fact that the multiplets are much more symmetric in CoRoT 105288363, while most peaks appear on he right side of the harmonics in V445 Lyr, resulting in positive values of f modulo $\rm{f}_0$.}
\label{echelle}
\end{figure}

One of the things the two stars have in common, is a secondary modulation that manifests itself as an additional frequency multiplet in the Fourier spectrum. The Blazhko effect normally leads to a regularly spaced frequency multiplet around the fundamental frequency and its harmonics, but in the two stars discussed here, a second regular pattern of peaks was intermingled with the classical multiplet. This secondary multiplet has higher amplitudes and is much clearer in V445 Lyr (where it also shows second order terms) than it is in CoRoT 105288363. Fig.~\ref{echelle} shows Echelle diagrams of the frequency pattern around the harmonics for both stars. Please note that these Echelle diagrams are not identical to those used in helioseismology, but in this case display the harmonics of the fundamental radial pulsation and serve as a practical tool to recognize patterns that repeat in every harmonic order.

The secondary modulation, however, was by no means sufficient to explain the variations in the modulation. After prewhitening both the classical multiplet and the secondary multiplet, a large number of partly unresolved peaks remained, indicating irregular behavior or a number of additional unresolved modulation periods. 

\subsection{Phasing between the modulation types}
One of the major findings during the analysis of CoRoT 105288363 was that the amplitude modulation and the phase modulation can change their relative contribution to the total modulation. While the amplitude variation during the four observed cycles became slightly weaker, the variation in the O-C diagram (which is related to the phase variation) was found to almost double its value by rising continuously from 20 min to 38 min. This had never been documented before. In V445 Lyr, an inspection of the two types of modulation yielded a similar result: amplitude modulation and phase modulation seem to change their strength independently from each other, and also do not keep their phasing relative to each other. Of eight observed Blazhko cycles in V445 Lyr, not one resembles another.

\subsection{Radial overtones and possible non-radial modes}
The second radial overtone has meanwhile been detected with low amplitudes in several stars \citep{poretti, ben, nemec}. Also in both stars discussed here it was found to be excited with frequency ratios $f_0/f_2$ of 0.590 and 0.585, respectively, and with average amplitudes of 0.5 and 1.4 mmag, respectively. Linear models show that this frequency region is well in the expected range for the second overtone. Interestingly, the amplitude of the second overtone $f_2$ was found to vary with time in V445 Lyr. An analysis of small subsets of the data revealed that it is only excited during the Blazhko minima, while it does not reach significant amplitude during the other phases. A temporal resonance with the other radial modes (which change their frequency during the Blazhko cycle) was proposed as an explanation.

\begin{figure}
\includegraphics[width=82mm]{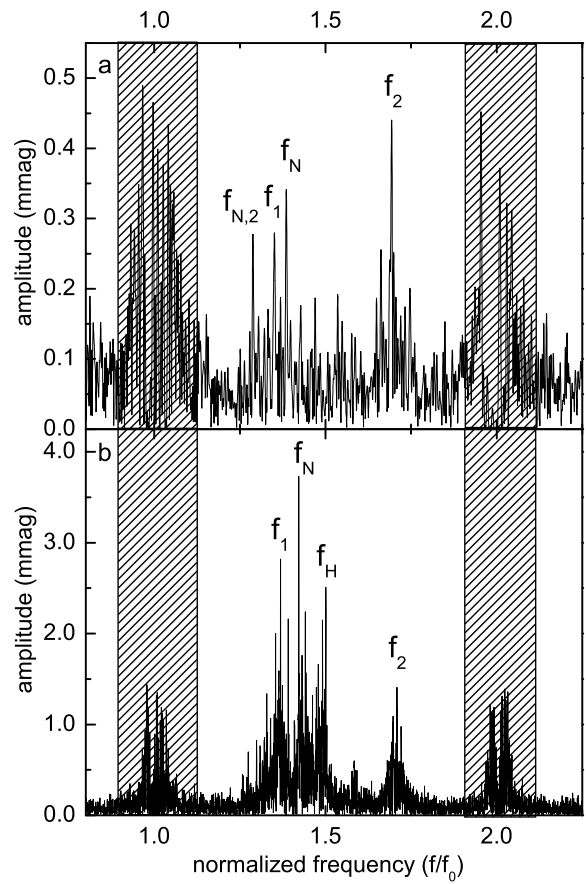} 
\caption{Normalized frequency spectra of CoRoT 105288363 (top panel) and V445 Lyr (bottom panel). The shaded boxes indicate the regions around $f_0$ and $2f_0$, where the Blazhko multiplet as well as the secondary multiplet have been prewhitened. In the region between the two harmonics, the additional peaks emerge clearly. Labels $f_1$ and $f_2$ designate the first and second overtone, respectively, $f_N$ and $f_{N,2}$ designate possible non-radial modes, while $f_H$ stands for the half-integer peak caused by the period doubling.}
\label{addmode}
\end{figure}

But there was much more to the two stars than radial overtones. Altogether four additional peaks were found in the region in between the harmonic orders in each of the two stars. In V445 Lyr, besides the above-mentioned second overtone, also the first overtone is probably present, as well as the frequencies caused by the so-called period-doubling phenomenon \citep{szabo}, and an additional mode that is most likely to be a non-radial mode. A revisit of the CoRoT 105288363 data set revealed a very similar frequency pattern. The effect of period doubling, however, is not detected in the CoRoT star, so that no half-integer frequency $f_H$ is present in the spectrum. Instead of that,  CoRoT 105288363 shows an additional frequency that might be a second non-radial mode. Fig. \ref{addmode} shows the frequency spectra of both stars. Spectra are normalized to the fundamental frequency so they are easy to compare.

The first possible detection of non-radial modes in an RR Lyrae star (unrelated to the components of a Blazhko modulation) was reported by \citet{gru} for the RRd star AQ Leo. This detection was also based on space photometry, in this case a 34~d long data set obtained by the MOST satellite. In this paper, however, we are dealing with RRab stars, for which non-radial modes have not been discovered before the advent of CoRoT and {\it Kepler}.

\section{Conclusions}
Based on the results discussed above, one can certainly see that the Blazhko effect is not necessarily a phenomenon that shows exact repetition from one cycle to the next. On the contrary: it can, at least in some cases, show strong signs of irregularity, with no identical two cycles in a set of eight observed cycles. This needs to be accounted for in models.
Also, the relative contribution of phase modulation and amplitude modulation can change in the sense that one of the two phenomena might weaken while at the same time the other one is getting stronger. The phasing between the two types of modulation is also observed to change, indicating that the two modulation types are not necessarily correlated.\\
Apart from the results on the Blazhko effect itself, one can see that RRab~Lyrae stars can have a zoo of additional frequencies excited with small amplitudes in addition to the fundamental mode, its harmonics, and the classical Blazhko frequency pattern. Among those frequencies, radial overtones as well as non-radial modes have been detected. Amplitudes and frequencies of those modes are not constant with time, indicating that resonances might be present, which could lead to transient excitation of the overtone modes. Additionally, the period doubling phenomenon, which was theoretically shown to be related to chaos (see also Molnar (2012), these proceedings), is present in one of the discussed stars.\\
Summing up, one can conclude that RR~Lyrae stars can be complicated objects with a multitude of phenomena and low-amplitude pulsation modes, and therefore should no longer be considered simple monoperiodic pulsators.\\
Continued space observations, as they are expected to be obtained by the extended {\it Kepler} mission, will increase the available time span, and will reveal if seemingly irregular behavior can be explained by a set of yet unresolved modulations periods. It will also open the possibility to detect irregularities in stars that seem to be regular on shorter time scales, and will show whether irregular Blazhko behaviour is wide-spread among RR Lyrae stars or restricted to a particular set of stars.

\acknowledgements
The author acknowledges support from the Austrian Science Fund (FWF), project number P19962-N16, and also gratefully
acknowledges the entire {\it Kepler} team, whose outstanding efforts have made these results possible. Funding for the {\it Kepler} discovery mission is provided by NASAÕs Science Mission Directorate.



\begin{thebibliography}{99}
  \bibitem[Balazs (1957)]{bala}Balazs, J.: 1957, Commun. Konkoly Obs., 42, 99 
  \bibitem[Benk\H{o} et al.(2010)]{ben}Benk\H{o}, J., Kolenberg, K., Szab\'o, R. et al.: 2010, MNRAS 409, 1585
  \bibitem[Blazhko(1907)]{blazhko}Blazhko, S.N.: 1907, AN, 175, 325
  \bibitem[Guggenberger et al.(2011)]{gug11}Guggenberger, E., Kolenberg, K., Chappelier, E. et al.: 2011, MNRAS 415, 1577
  \bibitem[Guggenberger et al.(2012)]{gug12}Guggenberger, E., Kolenberg, K., Nemec J.M. et al.: 2012, MNRAS 424, 649
  \bibitem[Gruberbauer et al.(2007)]{gru}Gruberbauer, M., Kolenberg, K., Rowe, J.F. et al.: 2007, MNRAS 379, 1498
  \bibitem[Kolenberg et al.(2010)]{kol}Kolenberg, K., Szab\'o R., Kurtz, D. et al.: 2010, ApJ 713, 198
  \bibitem[Jerzykiewicz \& Wenzel(1977)]{jer}Jerzykiewicz, M  \& Wenzel, W.: 1977, AcA 27, 35
  \bibitem[Jurcsik et al.(2009)]{jur}Jurcsik, J., S\'odor, \'A, Szeidl, B. et al.: 2009, MNRAS 400, 1006   
  \bibitem[Nemec et al.(2011)]{nemec}Nemec, J.M., Smolec, R., Benk\H{o}, J. M. et al.: 2011, MNRAS 417, 1022
  \bibitem[Poretti et al.(2010)]{poretti}Poretti, E., Papar\'o, M., Deleuil, M. et al.: 2010, A \& A 520, A108  
  \bibitem[Schwarzschild(1940)]{schw}Schwarzschild, M.: 1940, HarCi 437, 1   
  \bibitem[S\'odor et al.(2011)]{sodor}S\'odor, \'A, Jurcsik, J., Szeidl, B. et al.: 2011, MNRAS 411, 1585
  \bibitem[Szab\'o et al.(2010)]{szabo}Szab\'o, R., Koll\'ath, Z., Moln\'ar, L. et al.: 2010, MNRAS 409, 1244 
  \bibitem[Szeidl(1976)]{szeidl}Szeidl, B.: 1976, in Fitch W.S., ed., Proc. IAU Colloq. 29, 133

\end{thebibliography}
\end{document}